\documentclass[aps,prd,superscriptaddress,preprint,tightenlines,nofootinbib]{revtex4}

\usepackage{graphicx}% Include figure files
\usepackage{dcolumn}% Align table columns on decimal point
\usepackage{bm}% bold math
\usepackage{epsfig}

\def \numthrees{(5.93\pm 0.10)\times 10^6}
\def \numtwos{(9.11\pm 0.14)\times 10^6}

\def \pdgonesdilep{2.48}
\def \errpdgonesdilep{0.05}
\def \pdgtwosdilep{2.03}
\def \errpdgtwosdilep{0.08}

\def \datthrch{5215}
\def \datthrneu{6584} 
\def \datthrtwo{4391} 
\def \dattwoch{26417}
\def \dattwoneu{38069} 
\def \datthrinc{184760}
\def \dattwoinc{824418}

\def \erdatthrch{72}
\def \erdatthrneu{274} 
\def \erdatthrtwo{207} 
\def \erdattwoch{163}
\def \erdattwoneu{727} 
\def \erdatthrinc{430}
\def \erdattwoinc{908}

\def \effthrch{39.7}
 
\def \efftwoch{32.0}
 
\def \effthrinc{69.9}
\def \efftwoinc{50.3}

\def \ereffthrch{0.1}
 \def\ereffthrinc{0.2}
\def \erefftwoch{0.1}

\def \erefftwoinc{0.1}

\def \brfthrch{4.47} 
\def \brfthrneu{2.24} 
\def \brfthrtwo{1.82} 
\def \brftwoch{18.26} 
\def \brftwoneu{8.43} 
\def \brfthrinc{4.46} 
\def \brftwoinc{17.99} 

\def \brfthrchavg{4.46}

\def \erbrfthrchavg{0.13}
\def \brftwochavg{18.02}
\def \erbrftwochavg{0.61}

\def \sterbrfthrch{0.06}
\def \erbrfthrneu{0.09}
\def \erbrfthrtwo{0.09}
\def \sterbrftwoch{0.11}
\def \erbrftwoneu{0.16}
\def \sterbrfthrinc{0.01}
\def \sterbrftwoinc{0.02}

\def \systerrbrfthrch{0.18}
\def \systerrbrfthrneu{0.11}
\def \systerrbrfthrtwo{0.12}
\def \systerrbrftwoch{0.81}
\def \systerrbrftwoneu{0.42}
\def \systerrbrfthrinc{0.14}
\def \systerrbrftwoinc{0.59}

\def \prd#1#2#3{{Phys. Rev. D} {\bf#1}, #2 (#3)}
\def \prl#1#2#3{{Phys. Rev. Lett.} {\bf#1}, #2 (#3)}
\def \plb#1#2#3{{Phys. Lett. B} {\bf #1}, #2 (#3)}
\def \jpg#1#2#3{{J. Phys. G} {\bf #1} #2 (#3)}
\def \npb#1#2#3{{Nucl. Phys. B} {\bf #1} #2 (#3)}
\def \npa#1#2#3{{Nucl. Phys. A} {\bf #1} #2 (#3)}
\def \nima#1#2#3{{Nucl. Instrum. Methods Phys. Res., Sect. A} {\bf #1} #2 (#3)}

\def \etal{{\it et\,al.}}

\newcommand{\plus}{\mbox{$^{+}$}}
\newcommand{\minus}{\mbox{$^{-}$}}

\newcommand{\goesto}{\mbox{$\rightarrow$}}

\newcommand{\electron}{\mbox{$e^{-}$}}			% \negatron
\newcommand{\eminus}{\electron}
\newcommand{\positron}{\mbox{$e^{+}$}}
\newcommand{\eplus}{\positron}
\newcommand{\lplus}{\mbox{$\ell^+$}}
\newcommand{\lminus}{\mbox{$\ell^-$}}
\newcommand{\dilep}{\mbox{$\lplus\lminus$}}

\newcommand{\Bbar}{\mbox{$\bar{B}$} }
\newcommand{\BBbar}{\mbox{$B\Bbar$}}  

\newcommand{\jpsi}{\mbox{J/$\psi$}}

\newcommand{\upsi}{\mbox{$\Upsilon(1S)$}}
\newcommand{\upsii}{\mbox{$\Upsilon(2S)$}}
\newcommand{\upsiii}{\mbox{$\Upsilon(3S)$}}
\newcommand{\upsiv}{\mbox{$\Upsilon(4S)$}}
\newcommand{\upsv}{\mbox{$\Upsilon(5S)$}}
\newcommand{\upsns}{\mbox{$\Upsilon(nS)$}}
\newcommand{\upsms}{\mbox{$\Upsilon(mS)$}}

\newcommand{\chib}{\mbox{$\chi_{b}(1P)$}}
\newcommand{\chibp}{\mbox{$\chi_{b}(2P)$}}

\newcommand{\chibpot}{\mbox{$\chi_{b1,2}(2P)$}}

\newcommand{\piz}{\mbox{$\pi$}^{0}}

\newcommand{\pipi}{\mbox{$\pi$}^{+}\mbox{$\pi$}^{-}}
\newcommand{\pizpiz}{\mbox{$\pi$}^{0}\mbox{$\pi$}^{0}}
\newcommand{\mev}{\mbox{ MeV}}

\newcommand{\mevcsq}{\mbox{ MeV/}c^2}
\newcommand{\gevcsq}{\mbox{ GeV/}c^2}

\newcommand{\jpc}{\mbox{$J^{PC}$}}

\newcommand{\ee}{\mbox{$e\plus e\minus$}}
\newcommand{\gamgam}{\mbox{$\gamma\gamma$}}
\newcommand{\mumu}{\mbox{$\mu\plus\mu\minus$}}


\begin{document}
\preprint{CLNS 08/2032}
\preprint{CLEO 08-15}   % For conference papers  % For conference papers
\title{\boldmath
Improved Measurement of Branching Fractions for $\pi\pi$ Transitions among $\upsns$ States
}
%-------- INSERT HERE ------------
% Your author list goes here  REMOVE EVERYTHING to END INSERT and
% replace with your authorlist (ask cleoac).
\author{S.~R.~Bhari}\altaffiliation{Present address:  Department of Physics, California State University, Fullerton, California 92834, USA}
\author{T.~K.~Pedlar}
\affiliation{Luther College, Decorah, Iowa 52101, USA}
\author{D.~Cronin-Hennessy}
\author{K.~Y.~Gao}
\author{J.~Hietala}
\author{Y.~Kubota}
\author{T.~Klein}
\author{B.~W.~Lang}
\author{R.~Poling}
\author{A.~W.~Scott}
\author{P.~Zweber}
\affiliation{University of Minnesota, Minneapolis, Minnesota 55455, USA}
% ...
\collaboration{CLEO Collaboration}
\noaffiliation
\date{September 6, 2008}
\begin{abstract} 
Using samples of
$\numthrees\;\upsiii$ decays and $\numtwos\;\upsii$ decays
collected with the CLEO detector, 
we  report improved
measurements of the branching fractions for the following five transitions:
${\mathcal{B}}(\upsiii\goesto\upsi\pipi)=(\brfthrchavg\pm\sterbrfthrinc\pm\erbrfthrchavg)\%,$
${\mathcal{B}}(\upsii\goesto\upsi\pipi)=(\brftwochavg\pm\sterbrftwoinc\pm\erbrftwochavg)\%,$
${\mathcal{B}}(\upsiii\goesto\upsi\pizpiz)=(\brfthrneu\pm\erbrfthrneu\pm\systerrbrfthrneu)\%,$
${\mathcal{B}}(\upsii\goesto\upsi\pizpiz)=(\brftwoneu\pm\erbrftwoneu\pm\systerrbrftwoneu)\%$ and 
${\mathcal{B}}(\upsiii\goesto\upsii\pizpiz)=(\brfthrtwo\pm\erbrfthrtwo\pm\systerrbrfthrtwo)\%.$
In each case the first uncertainty reported is statistical, while the second is systematic.  
\end{abstract}
\maketitle
\begin{figure}
\includegraphics[width=\linewidth]{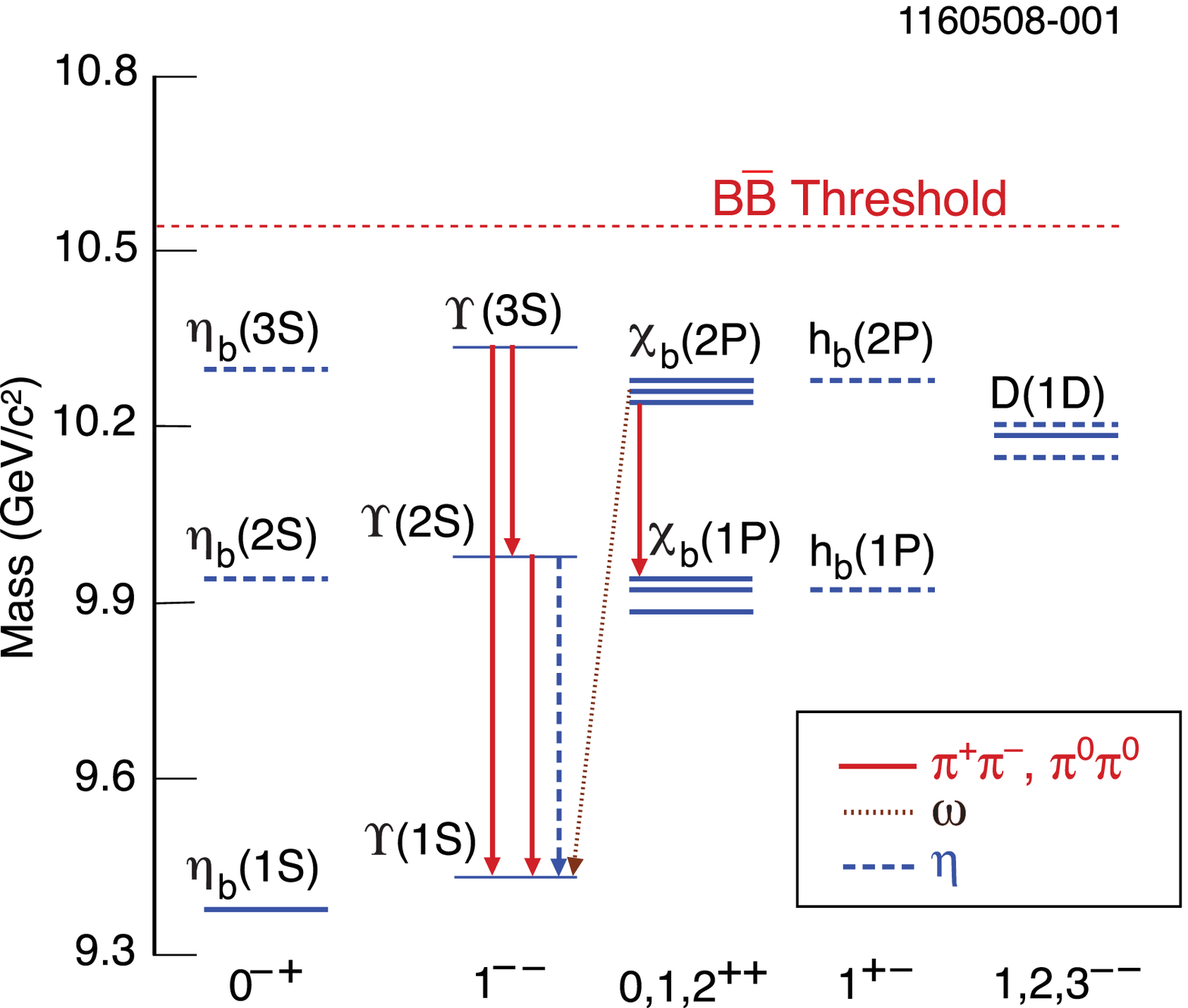}
\caption{The spectrum of bottomonium states below 
$\BBbar$ 
threshold for different $\jpc$ combinations. 
Established bottomonium states are indicated by the solid horizontal
lines, while those indicated by dashed horizontal lines have never been observed.
Arrows connecting various bottomonium states represent hadronic 
transitions that have been observed. \label{bbspec} 
}
\end{figure}

Hadronic transitions among heavy quarkonium states provide an excellent testing ground
for non-perturbative Quantum Chromodynamics (QCD)~\cite{kqrr}.   They are generally 
understood to proceed by the emission and hadronization of low momentum gluons~\cite{gottfried},  
and their investigation is one of few possible laboratories for the study of the low-$q^2$ hadronization process.
The study of such transitions in the bottomonium ($\bar{b}b$) system is particularly advantageous because 
of the non-relativistic nature of the system and the richness of the spectrum of states below open-bottom threshold.  (See Figure \ref{bbspec}.)

For the first 22 years after the observation of hadronic transitions among bottomonium states by  
LENA~\cite{lena} and CUSB~\cite{cusb}, only the six $\pi\pi$ transitions among
the vector $\Upsilon(nS)$ bottomonia were known.  CLEO has recently observed three other examples of 
hadronic transitions in bottomonium:  
$\chibpot\goesto\omega\upsi$~\cite{omegacbx},
$\chibp\goesto\pi\pi\chib$~\cite{chibpaper}
and
$\upsii\goesto\eta\upsi$~\cite{etapaper}.  Very recently, the BaBar Collaboration has reported new
measurements of several hadronic transitions in the bottomonium system using bottomonium states
produced in ISR while running at the $\upsiv$ resonance~\cite{babarpaper}. 

In this Article we report improved measurements of the branching fractions for $\pi\pi$ transitions among the vector
states of the bottomonium system.  Dipion transitions from $\upsiii$ to the lower vector states ($\upsii{%
,\;\upsi}$) and from $\upsii$ to $\upsi$ have been of interest ever since their first observation in 
1982~\cite{expt3s}.  There has recently been a resurgence of interest in dipion transitions following the 
observation by Belle~\cite{belleups} and BaBar~\cite{babarups} of 
dipion transitions from the bottomonium resonances $\upsiv$ and $\upsv$ to lower bottomonium states,  of 
the new state Y(4260)~\cite{y4260} to $\jpsi$, and also the 
observation by BES and CLEO of similar transitions of $\psi(3770)$~\cite{bespsi3770,cleopsi3770} .  
Additional motivation to update measurements of 
the branching fractions for dipion transitions among bottomonium states 
below open-bottom threshold is presented by the prospects of using $\upsiii,\upsii\goesto\pi\pi\upsi$ as a clean
source of tagged $\upsi$ to study exclusive $\upsi$ decays, including searches for invisible decay modes~\cite{invis}.

A well-known feature of the $\pi\pi$ transitions in bottomonium is that the
invariant mass of the dipion system in 
$\upsiii\goesto\pi\pi\upsi$ differs greatly from that produced in 
other known dipion transitions in bottomonium and in charmonium~\cite{expt3s}.   Theoretical interest
in these invariant mass distributions has been substantial, and several attempts to describe them have
been made since the first observation of dipion transitions~\cite{theory}. Analysis of 
the dipion invariant mass shapes in transitions from bottomonium states above
open-bottom threshold show similar interesting features~\cite{belleups,babarups}.  
A detailed analysis of the dipion invariant mass shapes including the extraction of the matrix elements for 
the transitions considered in this Article was performed using the most recent CLEO data and 
appears in Ref.~\cite{pappascbx}.   The results of that matrix element determination are 
used in the present analysis to properly determine the detection efficiency.  

%\section{Data Sample}

For this work, data samples were collected using the CLEO III~\cite{cleo3det} detector at the Cornell Electron Storage Ring.  
These samples included $\numthrees\;\upsiii$ collected on the $\upsiii$ resonance, at $\sqrt{s}=10.355$ GeV
and $\numtwos\;\upsii$ decays collected on the $\upsii$ resonance, at $\sqrt{s}=10.025$ GeV.
Charged particle tracking is done by a 47-layer drift 
chamber and a four-layer silicon 
tracker immersed in a 1.5~T solenoidal magnetic field.
Photons are detected using an electromagnetic calorimeter
consisting of 7784 CsI(Tl) crystals in a 
projective barrel geometry.

In this analysis, we study the transitions both inclusively (in which case we detect
only the pair of charged pions) and exclusively (in which case we detect, in addition
to the charged or neutral pair of pions, the decay of the daughter $\upsns$ state to either
$\mumu$ or $\ee$).  In each case, the 
primary quantity used to identify our observation of the dipion transitions of
interest is the mass recoiling against the 
dipion system.  This may be most
simply defined in terms of the formula
$
%\begin{equation}
M_{\mathrm{recoil}}=
\sqrt{(E_{\mathrm{cm}} - E_{\pi\pi} )^2
-(\mathbf{p}_{\pi\pi})^2}$, where $E_{cm}$ is the energy in the center-of-mass system, 
$E_{\pi\pi}$ and $\mathbf{p}_{\pi\pi}$ are the energy and three-momentum
of the dipion system, respectively.  
%\end{equation}
For a dipion system produced in the transition 
from $\upsns$  to $\upsi$ ($\upsii$), $M_{recoil}$ will be equal,  within detector resolution, to the
$\upsi$ ($\upsii$) mass, 9.460 (10.023) $\gevcsq$~\cite{pdg}.  Randomly selected pion pairs from hadronic
events do not peak at all, but form a smooth combinatoric background.

In the inclusive analysis, we select two charged tracks
that originate within 5~cm in the beam direction and 5~mm in the transverse direction 
from the center of the interaction region.  The vertex requirements drastically 
reduce the likelihood that the $\pi$ candidate tracks were produced in interactions between the $\eplus$ or
$\eminus$ beams and the beampipe, residual gas, or other material.
Monte Carlo studies show that low momentum kaon or 
muon pairs arising from generic $\upsns$ or QED processes do not produce peaking backgrounds
in the recoil mass spectrum.  Therefore, we did not require positive identification of these 
tracks as pions.   The mass recoiling against 
the two tracks, assumed to be pions, was required to be greater than $9.0\gevcsq$.  

In the exclusive analysis we require events to have two high momentum 
tracks with an invariant mass of $9.2-9.7\gevcsq$, consistent 
with $M(\upsi)$, or greater than $9.9\gevcsq$, consistent with $M(\upsii)$.  
These tracks must have an angle with respect to the beam direction, $\theta_{\ell}$, which satisfies $|\cos\theta_{\ell}|<0.82$,
a region in which the acceptance is relatively uniform.
We apply no further track quality criteria for the lepton candidates, 
and we do not attempt to distinguish to which dilepton final state 
(electron or muon) the $\upsi$ candidate has decayed.  Since in the exclusive analysis we reconstruct
the full event, we require the sum of energies of all final state particles to be greater than 
$\sqrt{s} - 200\mev (-300\mev)$) for the $\upsiii$ ($\upsii$) analyses.  The more stringent requirement on the
$\upsiii$ energy conservation was necessary to remove possible contamination in the signal sample due to 
cascades through the $\upsii$.   

In the charged exclusive case, we require that in addition to the dilepton candidate, events have a pair of low momentum tracks that
satisfy the same requirements as the pion candidates in the inclusive analysis.
The dilepton invariant mass requirement alone provides a nearly background-free sample,
and imposition of additional criteria in order to identify the tracks as leptons only leads to larger 
systematic uncertainties and reduced signal efficiency without much improvement in  
signal quality.  An additional requirement is imposed
to remove radiative Bhabha events ($\ee\goesto\ee\gamma$) in which the $\gamma$ converts in the inner
material of the detector or the beampipe, producing an $\ee$ pair that can fake the transition $\pipi$.  In such
events a small angle between momentum of the conversion pair and one of the two
high-momentum leptons is favored.  Essentially all of this background is removed by 
the requirement that this
angle be greater than 0.15 radians. Finally, in 
addition to the four tracks (two $\pi^{\pm}$ candidates, two $\ell^{\pm}$ candidates) we allow
events to have one additional track, which prevents the loss of otherwise good events due to 
failures in pattern recognition or to other spurious track candidates.   Monte Carlo (MC) studies show that this allowance
does not contribute any peaking background in the region of interest. 

In the neutral dipion analysis, we require that in addition to the dilepton candidate, events contain four or five showers
in the calorimeter.  
Each of these showers must have an energy of at least $50\mev$ and have angles relative to the beam axis such that
$|\cos(\theta_{\gamma})| \leq 0.804$, where $\gamma$ reconstruction is best.  None
of these showers may be matched to either charged track in the event.  Showers satisfying these criteria are then paired to produce
$\piz$ candidates.  Pairs with invariant masses within $50\mevcsq$ of the nominal $\piz$ mass must further satisfy 
the requirement that the mass pull - the normalized deviation from the nominal $\piz$ mass, 
$(M_{\gamgam}-M_{\piz})/\sigma_{\gamgam}$ -  be in the range (-4.0,2.5) for $\upsns\goesto\upsi\pizpiz$ 
and (-5.0,3.0) for $\upsiii\goesto\pizpiz\upsii$.  The pair-mass resolution, $\sigma_{\gamgam}$, which is calculated event
by event based on the calorimeter energy deposits, is typically between 5 and 7 $\mevcsq$. The $\piz$ candidates satisfying 
this condition are then subject to a mass-constrained fit in order to improve the $\piz\piz$ recoil mass resolution. 

We have also allowed one additional spurious shower to be present in the calorimeter, in order that we do
not needlessly remove events due to spurious signals in the calorimeter.
Because we allow up to five showers in the calorimeter it is possible for an event to have more 
than two combinations of showers that
satisfy the $\piz$ candidate requirements.  In such cases,
the combination of $\piz$ pairings having the smallest sum 
of squared mass pulls.

In order to evaluate detector acceptances and efficiencies and to study backgrounds to the signal processes,
MC samples were generated for several different event types.
Generic $\upsiii$ and $\upsii$  decays and  
continuum processes (such as $\ee\goesto\tau\tau$) 
at center of mass energies equal to the masses of the two states
were simulated using the routine {\textsc{QQ}}~\cite{qq}. 
MC samples were also generated for the signal dipion transitions and for individual 
background channels using {\textsc{EvtGen}}~\cite{evtgen}.  Each sample was then
passed through a {\textsc{GEANT}}-based~\cite{geant} detector simulation.  
Generic and continuum MC samples contained approximately one and five times the 
actual integrated luminosity taken at each of the resonances, respectively.   

For the study of acceptance and efficiency, separate signal MC samples for
 inclusive and exclusive analyses were created.  In order to take advantage of the matrix element analysis
 previously performed by CLEO~\cite{pappascbx}, the signal MC samples were generated 
 according to phase space and then weighted according to the square of the matrix elements. 
In the exclusive analyses, a further weighting factor of  $1+\cos^2\theta^*_{\ell}$ (where
$\theta^*_{\ell}$ is the lepton angle relative to the beamline in the rest frame of the
daughter $\Upsilon$) was applied.  This assumes a negligible $D$-wave component in the 
$\pi\pi$ transition.
For the inclusive analyses,
200,000 events were generated for each of $\upsiii$ and $\upsii$ decaying by
$\pipi\upsi$, where $\upsi$ was decayed generically. 
For the exclusive analyses,
500,000 events were generated for each of the five transitions, 
with the daughter $\upsi$ or $\upsii$  decaying equally to $\ee$ and $\mumu$.

\begin{figure}[t]
\includegraphics[width=\linewidth]{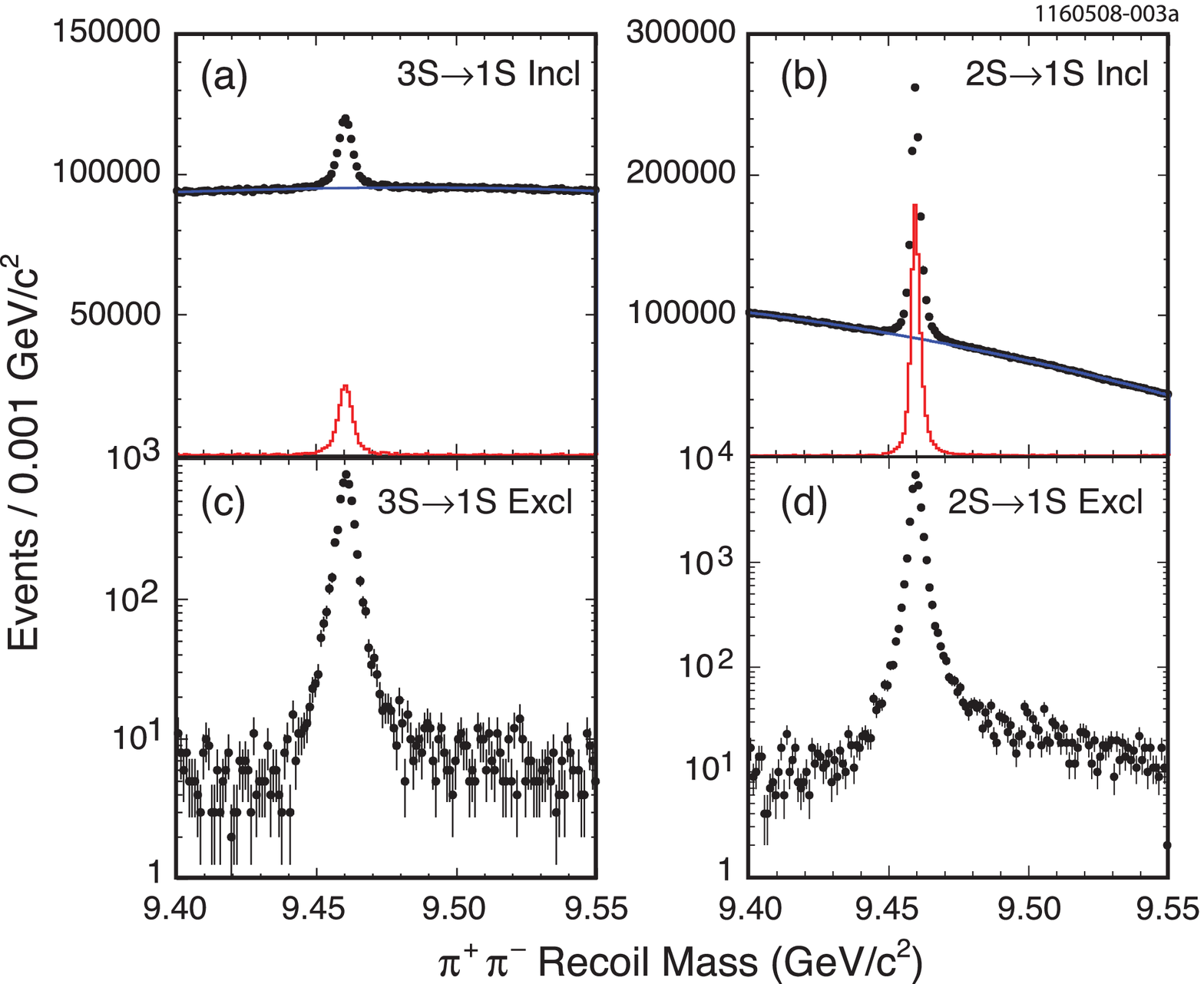}
\caption{$\pipi$ recoil mass distributions for the four charged dipion transition analyses,
(a) $\upsiii\goesto\pipi\upsi$ inclusive, 
(b) $\upsii\goesto\pipi\upsi$ inclusive,
(c) $\upsiii\goesto\pipi\upsi$ exclusive and
(d) $\upsii\goesto\pipi\upsi$ exclusive. 
The data are represented by symbols with uncertainty, while the 
histograms overlaid in each inclusive plot represents background-subtracted data.  Note the logarithmic scale for the exclusive 
analyses.\label{fig2}}
\end{figure}
\begin{figure}[t]
\includegraphics[width=\linewidth]{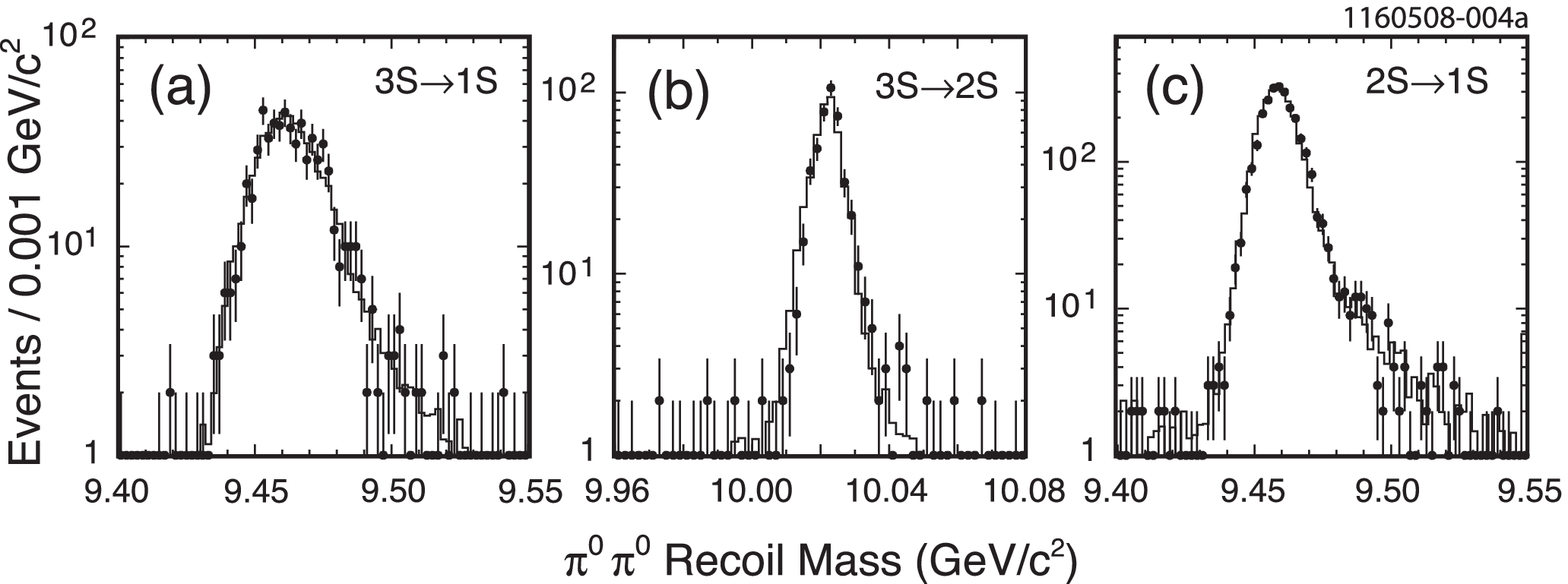}
\caption{$\pizpiz$ recoil mass distributions in the neutral dipion transition analyses, 
(a) $\upsiii\goesto\pizpiz\upsi$, (b) $\upsiii\goesto\pizpiz\upsii$ and (c) $\upsii\goesto\pizpiz\upsi.$  
The data are represented by the symbols with errors, while the histograms represent the result of the
fit using the MC shape and a linear background function.  Note the logarithmic scale.\label{fig3}}
\end{figure}

In all analyses, the distribution of the mass recoiling against the $\pi\pi$ pair for accepted data and MC events
is used to evaluate signal yield and efficiency, respectively.   Recoil mass distributions for all the transitions observed in our data
are shown in  Figs.~\ref{fig2} and~\ref{fig3}.

For the inclusive analysis, there is a large combinatoric background
because the analysis involves only combining pairs of charged pions, 
which are prolifically produced in $\Upsilon$ decays.  
The background due to these is smooth and has been
fitted to a third-order polynomial, and then subtracted in order to evaluate the yield for the signal process of
interest.  A study of continuum data taken below the $\upsiii$
and $\upsii$ resonances, and both continuum and 
generic $\upsiii$ and $\upsii$ MC simulations (which include decays of daughter $\upsi$ states) 
reveal no peaking background in $\pipi$ recoil mass.  

For the exclusive charged analysis, background events arise either from non-$\pi\pi$ 
hadronic transitions to the daughter $\Upsilon$ state in which the hadrons 
involved in the transition fake the signal due to poor reconstruction or noise in the detectors, or 
from $udsc$ quark pair production from the continuum.  Backgrounds arising from these
process should be negligible, because of the lack of significant non-$\pi\pi$ hadronic transitions in
the bottomonium system, and because of the required detection
of a high mass dilepton.  Continuum data taken below $\upsiii$ and $\upsii$ resonances and both continuum 
and generic MC simulations (with simulated $\pipi$ transitions removed) have been analyzed, 
and this expectation is confirmed. 

For the neutral analysis, there are also a number of possible background processes to consider.  One such 
process  is the decay chain $\upsiii\goesto\gamma\chibp$, $\chibp\goesto\gamma\upsii$, $\upsii\goesto\gamma\chib$,
$\chib\goesto\gamma\upsi$, $\upsi\goesto\dilep$.  The resulting $4\gamma + \dilep$ 
final state has an overall branching fraction of approximately $0.2\%$, while the branching fractions of
the signal processes are of order several percent~\cite{pdg}.  Such a process is furthermore unlikely to 
produce a fake signal because the four photons produced in the cascade would need to combine 
to produce two good $\piz$ candidates.
Similarly, the two-photon cascades $\upsiii\goesto\gamma\chi_b(2P,1P)$, followed by $\chi_b(2P,1P)\goesto\gamma\upsi$ 
and $\upsi\goesto\dilep$ could potentially fake the $\pizpiz$ signal with the addition of two spurious clusters in
the calorimeter.  To evaluate these small but possibly pernicious backgrounds, 
special MC samples, each of 50,000 events, were produced for both the
four and two photon cascade processes described above.  Analysis of
these samples show that such processes contribute negligible background.
Finally, transitions involving $\eta$ could as well, but because
of the tiny branching fraction for $\eta$ transitions~\cite{etapaper}, they 
have been assumed to produce no background.

To determine the branching fractions for the charged exclusive analyses, we used a cut-and-count method in which
the background underneath the peak in the recoil mass distribution was subtracted by means
of 20 MeV wide sidebands separated by 15 MeV from the 40 MeV wide signal region. 
We analyzed the inclusive recoil mass distributions similarly,
first subtracting the smooth combinatoric background followed by cut-and-count using the same signal and sidebands
as in the exclusive analysis.  This method minimizes the systematic uncertainties
associated with the MC signal shape, and corrects for any inaccuracies resulting from the
use of the background function in the vicinity of the signal peak.  Finally, from the resulting
data recoil mass histograms, we obtain the data yield, and from the signal MC, we obtain the efficiency
by dividing the MC yield by the amount of MC generated. 

For the neutral analyses we instead obtained the efficiency-corrected data yield directly by
fitting the data recoil mass histogram using the recoil mass histogram shape found from neutral  
signal MC samples, allowing for a first degree polynomial background contribution. 

From the data yields and MC efficiencies, we then calculate
the branching fraction, using
\begin{equation}
{\mathcal{B}} (\upsns\goesto\upsms\pi\pi) = \frac{N_{\mathrm{exc}}/\epsilon_{\mathrm{exc}}}{[N(\upsns){2\mathcal{B}}(\upsms\goesto\mumu)]},
\end{equation}
in the case of the exclusive analyses
and 
\begin{equation}
{\mathcal{B}} (\upsns\goesto\upsi\pipi) = \frac{N_{\mathrm{inc}}/\epsilon_{\mathrm{inc}}}{[N(\upsns)]},
\end{equation}
in the case of inclusive studies.  

The branching fractions obtained are
summarized in Tables~\ref{tab1} and~\ref{tab2}.
In order to evaluate the exclusively-measured branching fraction, 
we have assumed lepton universality.  
We have therefore used 
twice the PDG average value of ${\cal{B}}(\upsi\goesto\mumu) = \pdgonesdilep\pm\errpdgonesdilep\%$
to normalize our results for the transitions terminating in $\upsi$,  
 and 
twice the recent CLEO measurement of ${\cal{B}}(\upsii\goesto\mumu) = \pdgtwosdilep\pm\errpdgtwosdilep\%$~\cite{istvan}
for the one terminating in $\upsii$.

\begin{table}[t]
\caption{Results of the branching fraction measurements for charged dipion transitions
$\upsiii\goesto\pipi\upsi$ and $\upsii\goesto\pipi\upsi$. 
 The first uncertainty in the branching fraction
is the statistical uncertainty, while the second is systematic.  The averages listed are
weighted averages of the inclusive and exclusive measurements which take into
account correlation of systematic uncertainties between them.\label{tab1} }
\label{chtable}
\begin{center}
\begin{tabular}{lccc}
\hline\hline
Analysis & Data Yield & Efficiency ($\%$) &${\cal{B}}$ ($\%$)\\ 
\hline
3S Excl. & $\datthrch \pm \erdatthrch$ & $\effthrch\pm\ereffthrch$ & $\brfthrch\pm\sterbrfthrch\pm\systerrbrfthrch$ \\
3S Incl.& $\datthrinc \pm \erdatthrinc$ & $\effthrinc\pm\ereffthrinc$ & $\brfthrinc\pm\sterbrfthrinc\pm\systerrbrfthrinc$ \\
\hline
Average & & & $\brfthrchavg\pm\sterbrfthrinc\pm\erbrfthrchavg$\\
\hline
2S Excl.& $\dattwoch \pm \erdattwoch$ & $\efftwoch\pm \erefftwoch$ & $\brftwoch\pm\sterbrftwoch\pm\systerrbrftwoch$ \\
2S Incl.& $\dattwoinc \pm \erdattwoinc$ & $\efftwoinc\pm \erefftwoinc$ & $\brftwoinc\pm\sterbrftwoinc\pm\systerrbrftwoinc$ \\
\hline
Average && &$\brftwochavg\pm\sterbrftwoinc\pm\erbrftwochavg$\\
\hline\hline
\end{tabular}%
\end{center}
\end{table}
\begin{table}[t]
\caption{Results of measurements of the branching fractions for neutral dipion transitions.
The middle column lists the efficiency-corrected data yield, which is obtained as described in the text.
The  statistical uncertainty presented accounts for both the data and finite MC statistics, and the second
uncertainty reflects the remaining systematic contributions.
\label{tab2}}
\begin{center}
\begin{tabular}{lcc}
\hline\hline
Analysis & Efficiency-corrected Yield & ${\cal{B}}$($\%$) \\ 
\hline
$3S\goesto 1S\pizpiz$ &$\datthrneu\pm \erdatthrneu$ & $\brfthrneu\pm\erbrfthrneu\pm\systerrbrfthrneu$ \\
$3S\goesto 2S\pizpiz$ &$\datthrtwo\pm \erdatthrtwo$ & $\brfthrtwo\pm\erbrfthrtwo\pm\systerrbrfthrtwo$ \\
$2S\goesto 1S\pizpiz$ &$\dattwoneu\pm \erdattwoneu$ & $\brftwoneu\pm\erbrftwoneu\pm\systerrbrftwoneu$ \\
\hline\hline
\end{tabular}
\end{center}
\end{table}

Systematic error contributions  are summarized in Table~\ref{tab3}. 
For event reconstruction in the exclusive analyses, 
a systematic uncertainty of $1.2\%$ per pair of charged pions and $3.2\%$ per pair of
neutral pions was assessed.   These systematic uncertainties were evaluated by comparing the 
ratio of event yield in the standard exclusive analysis to the yield obtained 
using an analysis that depends on the reconstruction of only one $\pi^{\pm}$ 
or $\piz$.    From this ratio a per-$\pi^{\pm}$ or $\piz$ uncertainty was obtained, and
doubled to give the relative uncertainty for finding the pair. 
For the exclusive analyses, a systematic 
uncertainty of $1.0\%$ per lepton pair was similarly obtained.
For the inclusive analyses, based on tracking studies in a variety of neutral 
and charged multiplicity environments, we conservatively assign a systematic uncertainty of 
$2.4\%$ per pair of charged pions.

\begin{table*}[t]
\caption{Summary of relative systematic uncertainties on the measurement of
the branching fractions, expressed in percent.   Systematic uncertainties for the inclusive 
and exclusive analyses of charged
dipion transitions are separately listed in the table.
\label{tab3}}
\begin{center}

\begin{tabular}{cccccccc}
\hline\hline
& \multicolumn{4}{c}{$\upsiii\goesto$} & \multicolumn{3}{c}{$\upsii\goesto$}\\
\hline
Contribution  &  $\upsi\pipi$& $\upsi\pipi$ & $\upsi\pizpiz$ & $\upsii\pizpiz$ &  $\upsi\pipi$ & $\upsi\pipi$ & $\upsi\pizpiz$\\
& Excl. & Incl. &  & & Excl. & Incl. & \\
\hline
$\pi^{\pm}$/$\piz$ & $1.2$ & $2.4$ & $3.2$ & $3.2$& $1.2$ & $2.4$ & $3.2$ \\
$ \ell$ Tracks & $1.0$ & N/A & $1.0$ & $1.0$& $1.0$ & N/A & $1.0$\\
Luminosity & $1.7$ & $1.7$ & $1.7$ & $1.7$& $1.5$ & $1.5$ & $1.5$\\
$\ell$ Type & $2.5$ & N/A & $2.5$ & $2.5$  & $2.5$ & N/A & $2.5$  \\
\hline
MC Modelling& $0.2$ & $0.4$ & $0.5$ & $2.2$& $2.3$ & $1.4$ & $0.2$\\
$ \ell \ell$ BR & $2.0$ & N/A & $2.0$ & $4.2$& $2.0$ & N/A & $2.0$ \\
Other Sources & $0.35$ & $0.8$ & $1.0$ & $1.0$& $0.1$ & $0.8$ & $1.0$\\
\hline 
Total & $4.0$ & $3.1$ & $5.1$ & $6.6$& $4.5$ & $3.3$ & $5.0$\\
\hline\hline
\end{tabular}

\end{center}
\end{table*}

For the $\upsiii$ analyses, a common relative uncertainty of $1.7\%$ due to the uncertainty
in the number of $\upsiii$ produced; for $\upsii$, the corresponding uncertainty was $1.5\%$.
This class of systematic uncertainty derives primarily from the uncertainty in the knowledge of
the integrated luminosity accumulated at each of the resonances.

Three sources of systematic uncertainty produce relatively large contributions in some cases. 
Uncertainties due to modelling of the dipion dynamics were studied by varying 
the MC weighting according to the uncertainties of the matrix elements reported in 
Ref.~\cite{pappascbx}, and by studying the resulting reproduction of 
the dipion invariant mass.   In the case of the $\upsiii\goesto\upsii\piz\piz$ and both
analyses of $\upsii\goesto\upsi\pipi$ estimated systematic 
contributions of $1.4-2.3\%$ were obtained.   For the other transitions,
the systematic uncertainty due to modelling was much smaller.  

The exclusive samples were all divided into  $\mumu$ and $\ee$ subsamples  for the purpose
of studying the difference between reconstruction of these two leptonic channels.  From the
difference in branching fractions obtained from the two subsamples, a relative 
systematic uncertainty of $2.5\%$
due to lepton type was obtained. The
exclusive analyses carry an additional systematic uncertainty from
the branching fractions for the decays of $\upsi$ and $\upsii$ to
dileptons.  

Uncertainties due to the choice of analysis requirements, MC statistics, side band range choices, etc., were
all much smaller in each case compared to the other systematic uncertainties.     
The overall relative systematic uncertainty is obtained by adding all contributions in quadrature.  
The complete array of systematic uncertainties for all seven analyses appears in Table~\ref{tab3}.  

We have measured the charged dipion branching fractions both inclusively and exclusively.  Our
final result for the measurement of each of these branching fractions is a weighted average of the
two independent results, which we have caluclated using a toy Monte Carlo method~\cite{lyons} in which we have 
properly accounted for the correlation between the various contributions to the systematic uncertainties that
are applied to the two analyses.  That is, the luminosity uncertainties on the exclusive and inclusive results 
are fully correlated, while the statistical uncertainties are uncorrelated, as are any uncertainties unique to either 
the exclusive or the inclusive analysis.  These average values are 
${\mathcal{B}}(\upsiii\goesto\upsi\pipi)=(\brfthrchavg\pm\sterbrfthrinc\pm\erbrfthrchavg)\%$ and
${\mathcal{B}}(\upsii\goesto\upsi\pipi)=(\brftwochavg\pm\sterbrftwoinc\pm\erbrftwochavg)\%.$

It is interesting to compare the branching fractions for the $\pizpiz$ transitions to those for the
corresponding $\pipi$ branching fractions that we have measured.  Isospin conservation requires that the square of the
matrix elements for the $\pizpiz$ transitions be half that of the $\pipi$ transitions.  Phase space
for the two types of transitions also differs slightly and modifies this expectation, 
such that the expected ratio ${\cal{B}}(\pizpiz\upsi)/{\cal{B}}(\pipi\upsi)$ 
is 0.53 for the transitions from $\upsii$ and 0.51 for transitions from $\upsiii$. 
Combining our neutral and charged results, and taking into proper account correlations and cancellations
among individual systematic errors, we obtain ratios of $0.462\pm 0.037$  for $\upsii$ transitions and
$0.501\pm 0.043$ for the transitions from $\upsiii$.  

%\section{Summary}
In summary, 
we have reported improved measurements of five of the six 
dipion transitions among the lower-lying bottomonium vector states 
$\upsiii$, $\upsii$ and $\upsi.$  Each of the measurements is more precise
than those made by any previous experiment, and also more precise than
the current PDG world average~\cite{pdg}.  

We gratefully acknowledge the effort of the CESR staff 
in providing us with
excellent luminosity and running conditions.
This work was supported by the 
A. P. Sloan Foundation, 
the National Science Foundation,
the U.S. Department of Energy,
the Natural Sciences and Engineering Research Council of 
Canada, and the U. K. Science and Technology Facilities Council.

\end{document}